\documentclass[amssymb,useAMS,prd,aps,amsmath,twocolumn,superscriptaddress,nofootinbib,preprintnumbers]{revtex4-1}
\usepackage{graphicx}
\usepackage{xcolor}
\newcommand{\remove}[1]{}
\usepackage{hyperref}

\begin{document}

\title{
Can Nonstandard Neutrino Interactions explain the XENON1T spectral excess?}
\author{Amir N.\ Khan}
\email{amir.khan@mpi-hd.mpg.de}
\affiliation{Max-Planck-Institut f\"{u}r Kernphysik, Postfach 103980, D-69029
Heidelberg, Germany}

\begin{abstract}
\noindent
We perform a constrained spectral fit analysis of the excess observed in the electron recoil energy spectrum by
XENON1T with neutrino magnetic moment, charge radius, neutrino millicharge and new light vector and scalar
mediators. 
Within the limits allowed by other laboratory experiments we find that the 
excess can be explained in the range$\ (2-4)\times 10^{-11}\mu
_{B} $ for the magnetic moment,$\ (2-6)\times 10^{-31}$cm$^{2}$ for the charge radius, $\ (1.7-2.3)\times 10^{-12}e$ for the millicharge and $(10-100)\ $keV masses of light mediators with couplings of $ 3.5\times 10^{-7}$ for vector/axial-vector, $1\times 10^{-6}$ and $4\times 10^{-6}$ for scalar and pseudo-scalar mediators respectively. Among all neutrino millicharge, magnetic moment and vector mediators fit better to the observed spectrum. We also derive constraints on all new physics parameters considered here.
\end{abstract}

\date{\today}
\pacs{xxxxx}
\maketitle

\section{Introduction}
\noindent
Recently, the XENON1T has observed an excess in the electron recoil energy
spectrum \cite{Aprile:2020tmw}.  
Above the known and expected Standard Model background, more events than expected were observed in the $(1-7)$ keV region of electron recoil. The large exposure and unprecedentedly low background rate of the experiment could indeed enable fundamental discoveries. Or enable the discovery of a new background source. 

Many phenomenological papers appeared immediately after the announcement of the collaboration \cite{ Boehm:2020ltd, AristizabalSierra:2020edu, Okada:2020evk, Bally:2020yid, Lindner:2020kko, Bally:2020yid,Alonso-Alvarez:2020cdv, Chala:2020pbn, Ge:2020jfn, Miranda:2020kwy,Amaral:2020tga,Benakli:2020vng, Chigusa:2020bgq, Li:2020naa, Baek:2020owl, Gao:2020wfr, Ko:2020gdg, An:2020tcg, McKeen:2020vpf, Bloch:2020uzh, Budnik:2020nwz,Smirnov:2020zwf, Jho:2020sku,Bramante:2020zos,Zu:2020idx,Fornal:2020npv}. This is one of them. We investigate the possibility that neutrinos possess new interactions that could modify the neutrino--electron scattering cross section at low energies accessible by XENON1T. 
Already in the XENON1T paper \cite{Aprile:2020tmw}, the possibility of an enhanced 
effective neutrino magnetic moment, although in comparison to \textit{only} Borexino limit, was analyzed, besides solar axions, bosonic dark matter and a possible new background source, tritium. Here, we take advantage of the bounds on the individual neutrino flavors reported by Borexino \cite{Borexino:2017fbd} and other experiments (See Table \ref{tabel1} below) on magnetic moment and other parameters. \par
\textit{}

With spectral analysis we explain this interesting excess with the help of nonstandard standard neutrino interactions (NSI) by considering bounds from reactor, accelerator, solar and COHERENT experiments \cite{Deniz:2009mu, Giunti:2015gga, Cadeddu:2020lky, Khan:2019cvi, Lindner:2018kjo, Arcadi:2018xdd, Dev:2019anc}. The possibilities we discuss are a neutrino magnetic moment, charge radius, millicharge, and the presence of light scalar and vector mediators that mediate neutrino-electron scattering. We will show how these new physics choices compete with each other in explaining the XENON1T observed spectral excess. 
\par Finally, we constrain all new physics parameters considered in this work using the XENON1T with 1 parameter at-a-time fits. We find that some of the bounds on neutrino magnetic moment and charge radius, millicharge and anapole moment of neutrinos are new and are in tension with the bounds from other laboratory experiments \cite{Giunti:2015gga, Cadeddu:2020lky, Khan:2019cvi}.  

\section{Expected NSI spectrum}

Here we recap the important formulas needed for our calculations of the
nonstandard neutrino interaction effects. We split them into neutrino
electromagnetic properties and light mediators in the neutrino-electron
interactions.

\textit{(1)}\ $\nu _{\alpha }-e$ \textit{\ weak and electromagnetic
scattering cross-sections- }\ The total differential cross section for the $%
\nu -e$ scattering is

\bigskip

\begin{equation}
\frac{d\sigma _{\nu _{\alpha }e}}{dE_{r}}=\left( \frac{d\sigma _{\nu
_{\alpha }e}}{dE_{r}}\right) _{SM}+\left( \frac{d\sigma _{\nu _{\alpha }e}}{%
dE_{r}}\right) _{_{MM}},
\end{equation}
where
\begin{equation}
\left( \frac{d\sigma _{\nu _{\alpha }e}}{dE_{r}}\right) _{SM}=\frac{%
2G_{F}^{2}m_{e}}{\pi }[g_{L}^{2}+g_{R}^{2}\left( 1-\frac{E_{r}}{E_{\nu }}%
\right) ^{2}-g_{L}g_{R}\frac{m_{e}E_{r}}{E_{\nu }^{2}}]
\end{equation}
is the standard electroweak cross section and 
\begin{equation}
\left( \frac{d\sigma _{\nu _{\alpha }e}}{dE_{r}}\right) _{_{MM}}=\frac{\pi
\alpha _{em}^{2} \mu _{\nu_{\alpha} }^{2}}{m_{e}^{2}}[\frac{1}{E_{r}}-\frac{1}{%
E_{\nu }}]
\end{equation}
is the neutrino magnetic moment cross section \citep{Fujikawa:1980yx, Vogel:1989iv,
Dvornikov:2003js, Dvornikov:2004sj}. Here, $G_{F}$\ is Fermi constant,$\ g_{L(R)}=(g_{V}\pm
g_{A})/2+1$ for $\nu _{e}$ and $g_{L(R)}=(g_{V}\pm g_{A})/2$ for $\nu _{\mu
} $ and $\nu _{\tau }$, $g_{V}=-1/2+\sin ^{2}\theta _{W}$, $g_{A}=-1/2$, $%
\mu _{\nu_{\alpha} }$ is the effective neutrino magnetic moment for each cross in units of Bohr
magneton $(\mu _{B}),$ $\alpha$ is the fine-structure constant, $m_{e}$
is the electron mass, $E_{\nu }$ is the incoming neutrino energy and $E_{r}$
is the electron recoil energy in the detector. We take $\sin ^{2}\theta _{W} =0.23867\pm 0.00016$ in the $\overline{\text{MS}}$ scheme \cite{Erler:2004in} with small radiative corrections, less than $2\%$, included.
Considering only the flavor conserving case for neutrino charge radius, neutrino millicharge  \citep{Fujikawa:1980yx, Vogel:1989iv,
Dvornikov:2003js, Dvornikov:2004sj} and anapole moment \cite{Rosado:1999yn}, we replace $g_{V}$ by $\widetilde{g}_{V}$, where%
\begin{equation}
\widetilde{g}_{V}=g_{V}\ +\frac{\sqrt{2}\pi \alpha }{G_{F}}\left(\frac{%
\left \langle r_{\nu _{\alpha }}^{2}\right \rangle }{3}-\frac{q_{\nu _{\alpha
}}}{m_{e}E_{r}}-\frac{a_{\nu _{\alpha }}}{18}\right).
\label{empro}
\end{equation}%
Here, $\left \langle r_{\nu _{\alpha }}^{2}\right \rangle$
and $a_{\nu _{\alpha }}$ are neutrino charge radius and anapole moment in units of $cm^{2}$ and  $q_{\nu _{\alpha}}$ is the neutrino fractional electric charge in units of
unit charge ``e''. Notice that among the three quantities neutrino millicharge appears to be more sensitive at the lower recoil and thus is expected to better fit the excess region.\par
\textit{(2) Light vector and scalar mediators- }$\ $\ For simplicity we
assume that new light mediators couple with equal strength to both neutrinos
and electron. For the vector (V)/axial-vector (A) type mediators, in the low momentum transfer
limit, we replace $g_{V/A}$ by $\widetilde{g}_{V/A}$, where \citep{Lindner:2018kjo, Arcadi:2018xdd}\ 
\begin{equation}
\widetilde{g}_{V/A} = g_{V/A}+\left( \frac{g_{Z^{^{\prime }}}^{2}}{\sqrt{2}%
G_{F}(2m_{e}E_{r}+m_{Z^{^{\prime }}}^2)}\right) ,
\end{equation}%
in the standard model cross section of eqn. 2, where $g_{Z^{^{\prime }}}$
is the coupling constant and $m_{Z^{^{\prime }}}$ is the mass of the vector/axial-vector
mediators. For scalar mediators we add their contribution to the standard
model cross section incoherently. The scalar (S) and pseudo-scalar (P) interaction cross sections  \citep{Cerdeno:2016sfi} therefore is%
\begin{equation}
\left( \frac{d\sigma _{\nu _{\alpha }e}}{dE_{r}}\right) _{_{S}}=\left( 
\frac{g_{S}^{4}}{4\pi (2m_{e}E_{r}+m_{S}^2)^{2}}\right) \frac{%
m_{e}^{2}E_{r}}{E_{\nu }^{2}},
\end{equation}%

\begin{equation}
\left( \frac{d\sigma _{\nu _{\alpha }e}}{dE_{r}}\right) _{_{P}}=\left( 
\frac{g_{P}^{4}}{8\pi (2m_{e}E_{r}+m_{P}^2)^{2}}\right) \frac{%
m_{e}E_{r}^2}{E_{\nu }^{2}},
\end{equation}%
where $g_{S }$ and $g_{P }$ are the scalar and pseudo-scalar coupling constant and $m_{S }$, $m_{P }$ are their masses, respectively.

Next we define the differential event rate in terms of the reconstructed
recoil energy $(E_{rec})$ in order to estimate the contribution of the
aforementioned new physics to the electron recoil spectrum. This can be
written as%
\begin{widetext}
\begin{equation}
\frac{dN}{dE_{rec}} = N_{e}\int_{E_{r}^{th}}^{E_{r}^{mx}}dE_{r} \int_{E_{\nu }^{mn}}^{E_{\nu }^{mx}}dE_{\nu
}\left( \frac{d\sigma _{\nu _{e}e}}{%
dE_{r}}\overline{P}_{ee}^{m} +\cos^2{\theta_{23}} \frac{d\sigma _{\nu _{\mu }e}}{dE_{r}}\overline{P}_{e\mu}^{m} +\sin^2{\theta_{23}} \frac{d\sigma _{\nu _{\tau}e}}{dE_{r}}\overline{P}_{e\tau}^{m}\right) \frac{d\phi }{dE_{\nu }}\epsilon (E_{rec})G(E_{rec},\
E_{r}),
\end{equation}
\end{widetext}
where $G(E_{r},\ E_{rec})$ is a normalized Gaussian smearing function to
account for the detector finite energy resolution with resolution power $%
\sigma (E_{rec})/E_{rec}=(0.3171/\sqrt{E_{rec}[\text{keV}]})+0.0015$ and $\epsilon
(E_{rec})$ is the detector efficiency both taken from \cite{Aprile:2020tmw, Aprile:2020yad},$%
\ d\phi /dE_{\nu }$ is the solar flux spectrum taken from \cite{Bahcall:2004mz} and $%
N_{e}$ is 0.65 ton-year exposure\cite{Aprile:2020tmw}. Here, $d\sigma _{v_{\alpha }e}/dE_{r}$ are cross sections
given in eqn. 2 above, $\overline{P}_{ee}^{m}$ $\ $ and $\overline{P}_{e\mu /\tau }^{m}$ are the oscillation length averaged
survival and conversion probabilities of solar $pp$ electron neutrino
including the small matter effects given as 
\begin{equation}
\overline{P}_{ee}^{m}=s_{13}^{4}+\frac{1}{2}c_{13}^{4}{}(1+\cos 2\theta _{12}^{m}\cos
2\theta _{12})
\end{equation}%
and $\overline{P}_{e\mu /\tau }^{m}=1-\overline{P}_{ee}^{m},$ where s$_{ij}$, c$_{ij}$ are mixing
angles in vacuum and $\theta _{12}^{m}$ is the matter effects induced mixing
angle taken from \cite{Lopes:2013nfa, Zyla:PDG2020}. We take values of oscillation parameters and their uncertainties from  \cite{Zyla:PDG2020} and for the analysis we consider only the normal ordering scheme. 
The integration limits are $E_{\nu
}^{mn}=(E_{r}+\sqrt{2m_{e}E_{r}+E_{r}^2})/2$ and $E_{\nu }^{mx}$ is the upper limit of each component of the PP-chain and CNO solar neutrinos considered here. We note pp neutrinos are the predominant contributors to energy range of interest here, while the other source have negligibly small effect on the total spectrum.
$E_{r}^{th}=1\ $keV is the detector
threshold and $E_{r}^{mx}=30$ keV is the maximum recoil energy for the region
of interest. 
\begin{figure}[tp]
\begin{center}
\includegraphics[width=3.5in, height=7.2in]{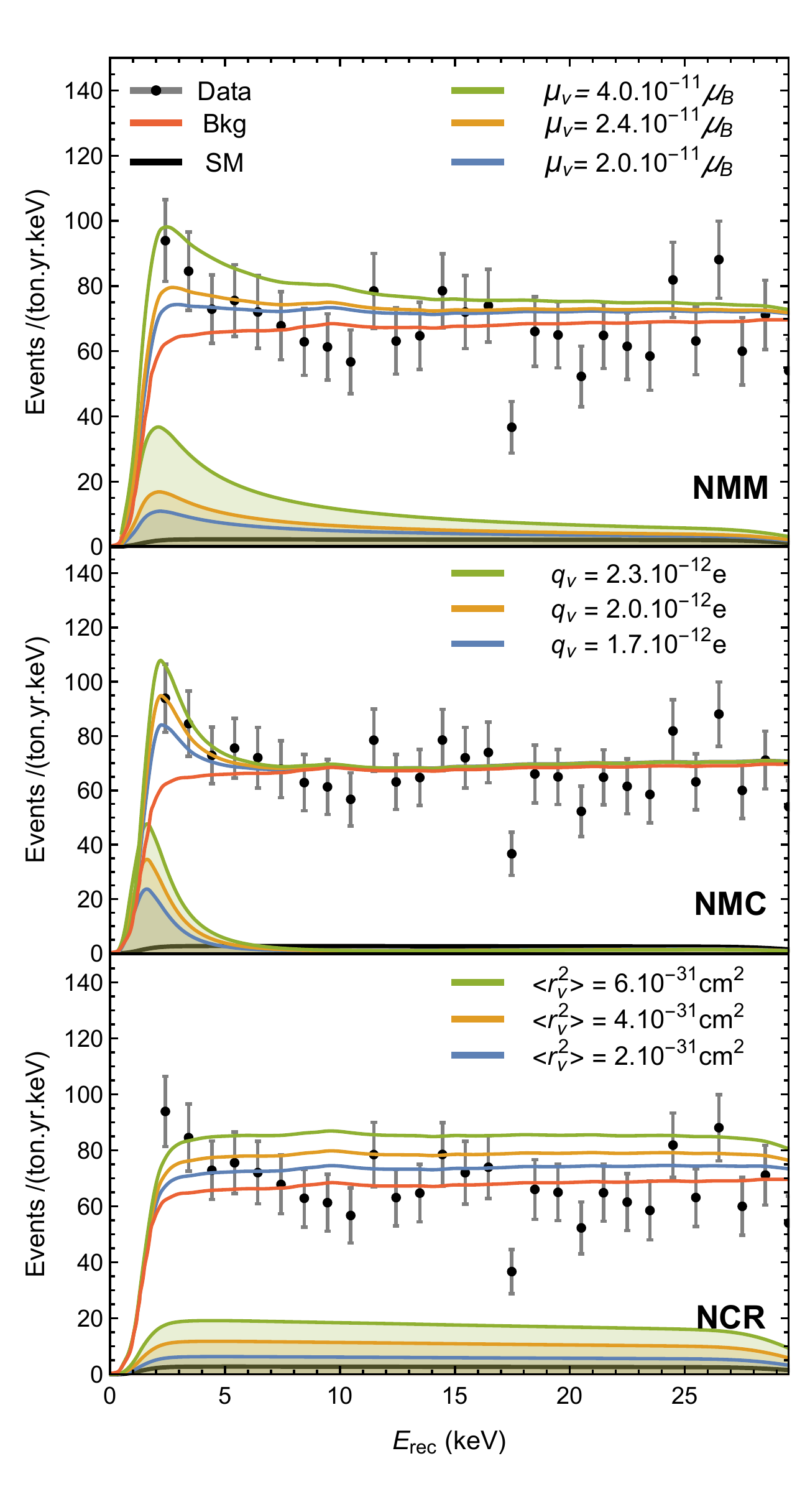}
\label{nuem}
\end{center}
\vspace*{-1.5cm}
\caption{{}\textbf{\ } Experimental data, backgrounds, SM energy spectrum and our
expected constrained fitted spectrum weighted with $1\sigma$ experimental uncertainty shown with each data points on neutrino magnetic moment (NMM) $(top)$, millicharge (NMC) $(middle)$ and charge radius (NCR) $(bottom)$. All fits were obtained with one parameter at a time. 
For each case, the central curve corresponds to the best fit while the outer two curves correspond to the constraint limits. Different graphs can be read off from the
legends inside each figure. The shaded regions in the bottoms correspond to
the expected without any backgrounds while the continuous line graphs
correspond to the expected+background. The data points and background
were taken from ref. \cite{Aprile:2020tmw}.}
\end{figure}

\begin{figure}[tp]
\begin{center}
\includegraphics[width=3.5in, height=7.2in]{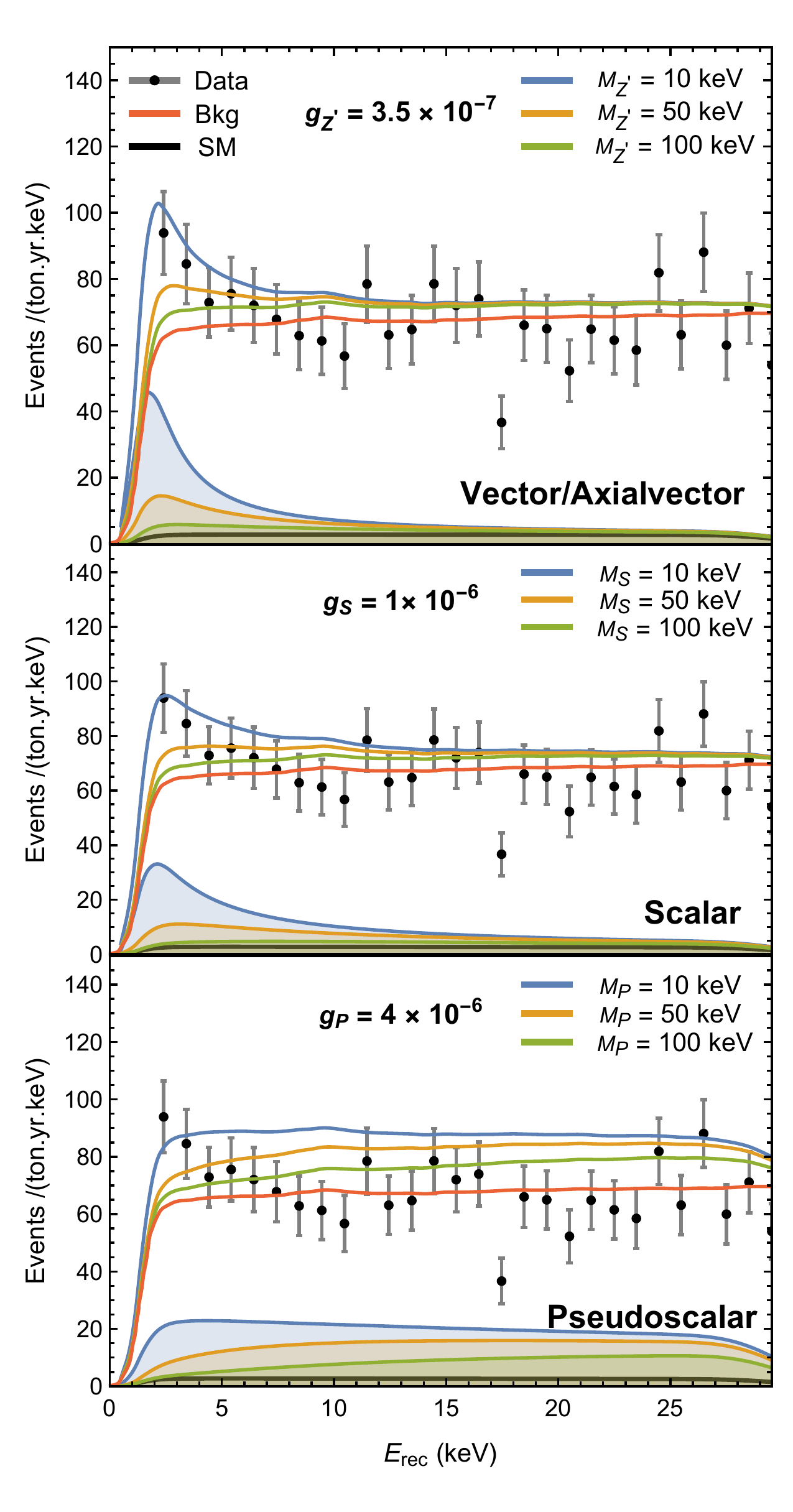}
\label{nuem}
\end{center}
\vspace*{-1.5cm}
\caption{{}\textbf{\ } Experimental data, backgrounds, SM energy spectrum and our
expected constrained fitted spectrum weighted with $1\sigma$ experimental uncertainty shown with each data points on vector/axial-vector $(top)$, scalar$(middle)$ and pseudoscalar $(bottom)$. All fits were obtained with one parameter at a time. 
For each case, the central curve corresponds to the best fit while the outer two curves correspond to the constraint limits. Different graphs can be read off from the
legends inside each figure. The shaded regions in the bottoms correspond to
the expected without any backgrounds while the continuous line graphs
correspond to the expected+background. The data points and background
were taken from ref. \cite{Aprile:2020tmw}.}
\end{figure}


\section{Results and Discussion}

Equipped with all the necessary formulas given above, we calculate the
differential event rate energy spectrum as function of $E_{rec}$ for the
standard model case and then for our new physics cases, that is,
neutrino magnetic moment, charge radius, millicharge, light vector and scalar
mediators. First, we made sure to reproduce the expected spectrum of fig. 1
of ref. \cite{Aprile:2020tmw}, corresponding to $\mu _{\nu }=7\times 10^{-11}$ $\mu
_{B}\ $including the true energy smearing and the efficiency of the detector
using our eqn. 8. Our SM\ expected energy spectrum shown in black in
both fig. 1 and 2 also agree very well with the expected spectrum given in
ref. \cite{shokley:slides}.
\par We note here that for all new physics that interferes with the SM, we normalize the expected new physics event distribution with our SM expectations. For the non-interfering new physics we simply add new physics as a signal above the given background ref. \cite{Aprile:2020tmw}, which already includes the SM expectations. Further, we note that our constraints are mainly dependent on the XENON1T neutrino magnetic moment result since we normalized our statistical model with several re-scaling parameters to exactly reproduce the XENON1T result on NMM. We discuss this in next section.

\subsection{Fits to the spectrum}
We include the neutrino magnetic moment, neutrino charge radius, millicharge, light
vector and scalar/pseudo-scalar mediators each one at-a-time and perform spectral fits to the observed spectrum within the constraints allowed by other laboratory experiments. For this purpose we collect all laboratory bounds on these parameters in Table I. While fitting to the observed spectrum, we put an additional constraint of $1\sigma$ experimental error. It is worthwhile to mention that in case of Borexino \cite{Borexino:2017fbd} we compare our fitting constraints with the bounds on the individual flavors, since, like Borexino, XENON1T receives fluxes from one survival and two conversion probabilities. This is clear from eq. 8 assuming the general case of non-maximal scheme of ``23'' sector. The results of these fits are shown in fig. $1$ and fig. $2$.

In Fig. $1$, the neutrino magnetic moment (top), the millicharge (middle) and charge radius (bottom) are shown. In each case the best fit spectrum corresponds to the central graph which is enveloped within the constraints applied while fitting. We note here that for the magnetic moment case one parameter at-a-time fitting gives best fit value of $~2.4\times 10^{-11}\mu _{B}$ which lies well within the interval obtained obtained by XENON1T. All the rest of the results shown in fig. 1 and fig. 2 were obtained by taking one parameter at-a-time. A general multi-parameter spectral fit with the same new physics or various types of new physics is beyond the scope of this work. We will turn to this issue in follow up work. \par
In Fig. 2, we show the best fit graphs (central) and their constrained limits (outer) for each case by fixing the coupling constants to the values given in each sub-figure and fitting the masses of the gauge boson.  In Fig. 2 vector/axial-vector  (top), scalar (middle) and pseudoscalar (bottom) cases are shown.  \par
Among the electromagnetic properties of neutrinos, the neutrino millicharge fits best to the excess region as well as to the over all spectrum, while the neutrino magnetic moment is the next leading and the charge radius has a relatively weaker fit. The neutrino charge radius has a flat increase across the whole spectrum. Among the light mediators, the vector/axial vector mediators are next leading while the scalar is comparatively better and the pseudoscalar has a relatively weaker fit. Like the charge radius, the pseudoscalar mediators have a flat increase across the whole spectrum. All of the parameters' best fits and constraints can be read off from Fig. 1 and 2.  


\subsection{Parameter constraints from XENON1T data}
Using XENON1T data we derive constraints on neutrino magnetic moment, charge radius, millicharge and anapole moment and all light mediators. For this purpose we define a modified $\chi^{2}$ function as follows 

\begin{equation}
\chi ^{2}=\underset{i,j}{\sum }a\left( \frac{%
b_{i}(dN/dE_{rec}+B(E_{rec}))_{th}^{j}-(dN/dE_{rec})_{obs}^{j}}{c_{i}\sigma
^{j}}\right)^2 
\label{chisq}
\end{equation}%
where $a$, $b_{i}$ and $c_{i}$ with $i=1,2$ are scaling factors introduced to exactly reproduce the XENON1T result on neutrino magnetic moment. We find $%
a=0.38,b_{1}=1,\ b_{2}=0.56, c_{1}=0.40,\ c_{2}=1.$ These factors
account for the uncertainties related to background and other systematics not given with the XENON1T data shown in fig. 7 of ref. \cite{Aprile:2020tmw}. The expression in the bracket
(.....)$_{th}^{j}$ represents the expected number of events in the $j-$th bin
while  bracket (...)$_{obs}^{j}$ is the observed number of events with $%
\sigma ^{j}$ as the uncertainty in the corresponding bin. We take data and errors from fig. 7 of ref. \cite{Aprile:2020tmw}. With the above $\chi ^{2}$ function, we reproduce the 90\% C.L. bound,  $[1.4,2.9] \times 10^{-11}\mu _{B}$, on neutrino magnetic moment and its $\Delta\chi^2$ distribution as shown with the thick black line graph corresponding to the $\nu_{e}$ flavor in the top left sub-figure of fig. 3.

\begin{figure*}[tp]
\begin{center}
\includegraphics[width=6.2in, height=3.7in]{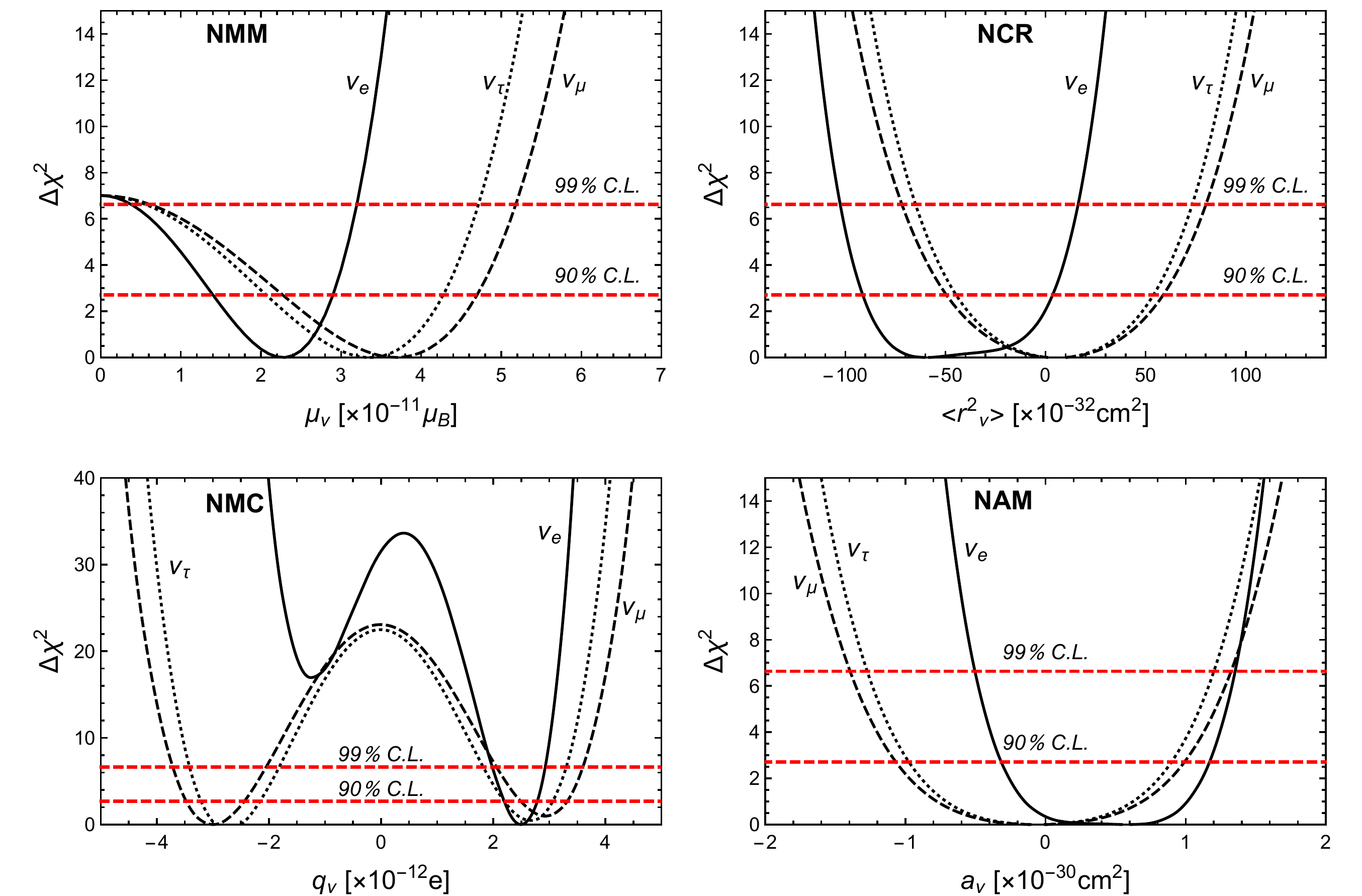}
\end{center}
\caption{{}\textbf{\ } $\Delta\chi^2$ distribution with 90\% and 99\% C.L. boundaries of neutrino magnetic moments (NMM), neutrino charge radius (NCR), neutrino millicharge (NMC) and neutrino anapole moment (NAM) of the three neutrino flavors from XENON1T data.}   
\label{Embounds}
\end{figure*}

\begin{figure*}[tp]
\begin{center}
\includegraphics[width=6.3in, height=3.7in]{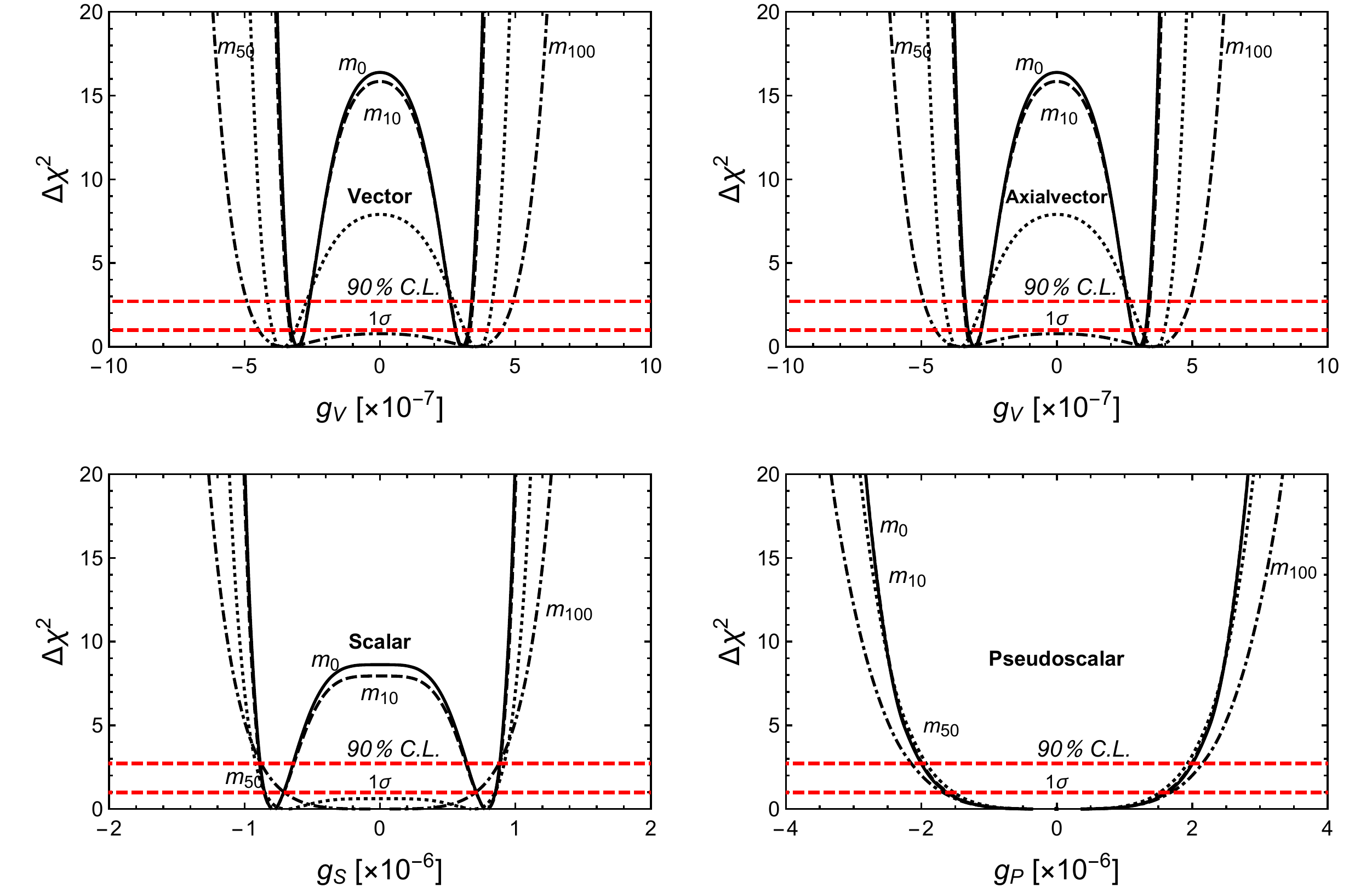}
\end{center}
\caption{{}\textbf{\ } $\Delta\chi^2$ distribution with $1\sigma$ and 90\% C.L. boundaries of flavor universal vector, axial-vector, scalar and pseudo-scalar mediator coupling constants using XENON1T data. The plot labels correspond to $m_{0}$ as massless mediators, $m_{10}$ as 10 keV, $m_{50}$ as 50 keV and $m_{100}$ as 100 keV.}   
\label{Lightmedi}
\end{figure*}

The results with $\Delta\chi^{2}$ distribution and the 90\% and 99\% C.L. projections for the electromagnetic parameters are shown in fig. \ref{Embounds}. Relying on reproducing the result for magnetic moment,  we use the same $\chi^{2}$ function of eq. \ref{chisq} to constrain all the rest of parameters. The 90\% C.L. bounds for the neutrino magnetic moment, charge radii, millicharge and anapole moments in flavor basis were extracted from fig. \ref{Embounds} and are shown in Table I. For comparison we also show bounds from other laboratory experiments. About Borexino, notice that they have not only reported the effective magnetic moment for the solar neutrinos, but also bounds on each neutrino flavor. See reference \cite{Borexino:2017fbd}.
For light mediators we fit the coupling constants for each case for four different choices of the masses: massless gauge bosons, 10 keV, 50 keV and 100 keV. The results from these fittings are shown in fig. \ref{Lightmedi} with $1\sigma$ and 90\% C.L. projection with the full parameter space.  \par
By generals inspection of fig. 3 and Table I, the constraints on the millicharge and magnetic moment are stronger than the existing bounds while the neutrino charge radii have weaker constraints. The constraints on the anapole moments were derived for first time in this analysis. \par

\begin{figure}[tp]
\begin{center}
\includegraphics[width=3in, height=5in]{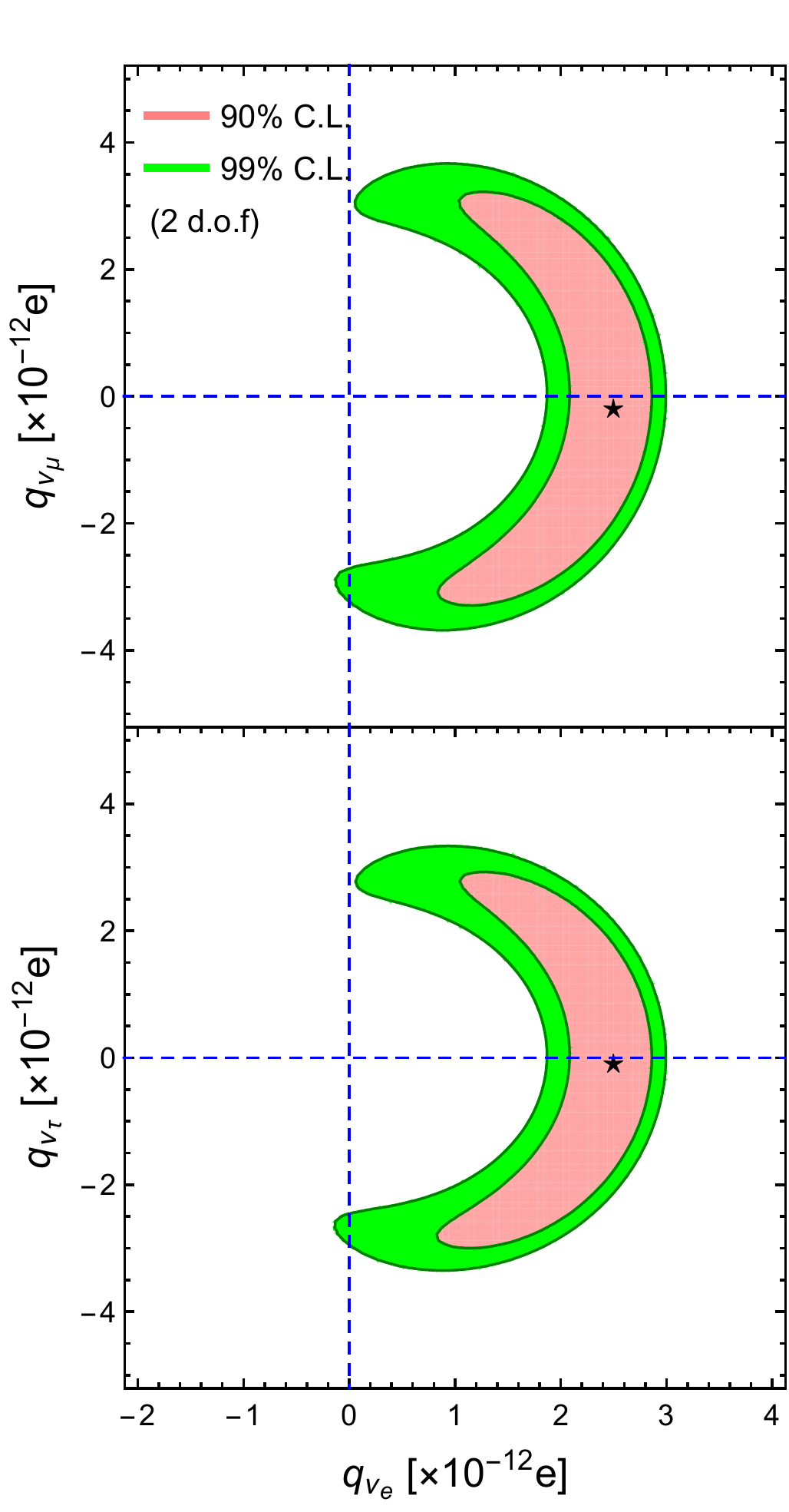}
\label{nuem}
\end{center}
\vspace*{-0.8 cm}
\caption{{}\textbf{\ } Two parameter $\Delta\chi^2$ contour in the plane of neutrino millicharges $q_{\nu_{e}}-q_{\nu_{\mu}}$ (Top) and $q_{\nu_{e}}-q_{\nu_{\tau}}$ (Bottom) at 90\% and 99\% C.L. See text for further details.}
\label{2d_NMC}
\end{figure}

As evident from the middle panel of fig. 1, the neutrino millicharge renders the best fit to the data among all the new physics choices. We further explore it with a two parameter fitting and the result is shown in fig. \ref{2d_NMC}. We show the millicharge contours in the $\nu_{e}$-$\nu_{\mu}$ and $\nu_{e}$-$\nu_{\tau}$ planes with the confidence level for 2 degrees of freedom. From fig. 5, it is clear that $\nu_e$ flavor completely excludes the negative valued region due to its large $\chi^2$ values for the negative values of NMC. This is more prominent from the NMC panel of fig. 3. On the other hand, the contour boundaries for the $\nu_{\mu}$ and $\nu_{\tau}$ flavors are symmetric about the SM axis in both fig. 3 and fig. 5. The positive region bias of $\nu_e$ is due to their larger solar flux in comparison to the other two flavors. As a result one can expect a larger deviation from SM through $\nu_{e}-e$ scattering in case of the XENON1T observed excess. We remind that the $\nu_e$ includes charge current contribution (via $W^{\pm}$) along with the neutral current (via Z) as for the $\nu_{\mu}$ and $\nu_{\tau}$ flavors. Taking intro account these factors, any new physics through electromagnetic interactions of neutrinos should be more enhanced for the $\nu_{e}$ flavor in the solar flux. This is very clear from fig. 3 and fig. 5 for NMC. In fig. 5, the best fit point for $(q_{e}, q_{\mu})$ corresponds to $(2.5, -0.2)\times 10^{-12}e$ while for $(q_{e}, q_{\tau})$, it corresponds to $(2.5, -0.1)\times 10^{-12}e$.

\begin{table*}[t]
\begin{center}
\begin{tabular}{c|c|c|c|c}
\hline \hline
Flavor & $\mu _{\nu }[\times 10^{-11}\mu _{B}]$ & $\left \langle r_{\nu
}^{2}\right \rangle \ [\times 10^{-32}$cm$^{2}]$ & $q_{v}\ [\times 10^{-12}e]$
& $a_{\nu }\ [\times 10^{-30}$cm$^{2}]$ \\ \hline
$\nu _{e}\ ($XENON1T$)$ & $[1.4,2.9]$ & $\ [-91,\ 4.1$ $]$ & $\  \ [2.2,\ 2.8$ 
$]\ $ & $\ [$\ $-0.32,1.2]$ \\ 
$\nu _{\mu }($XENON1T$)$ & $[2.26, 4.7]$ & $\ [-50,\ 59$ $]$ & $[-3.5,\ 3.3$ $%
]\ $ & $[-1.0,1.0]$ \\ 
$\nu _{\tau }($XENON1T$)$ & $[2.1, 4.3]$ & $\ [-44,\ 54$ $]$ & $[-3.2,\ 3.0$ 
$]\ $ & $[-0.98,0.91]$ \\ \hline \hline
$\nu _{e}\ ($Others$)$ & \multicolumn{1}{|l|}{$%
\begin{array}{l}
\leq 3.9\  \text{(Borexino)}\  \\ 
\multicolumn{1}{c}{\leq 110\  \text{(LAMPF)}\ } \\ 
\leq 11\  \text{(Super-K)} \\ 
\multicolumn{1}{c}{\leq 7.4\  \text{(TEXONO)}} \\ 
\multicolumn{1}{c}{\leq 2.9\  \text{(GEMMA)}}%
\end{array}%
$} & $\  \ 
\begin{array}{l}
\lbrack 0.82,\ 1.27]\  \text{(Solar)} \\ 
\lbrack -5.94,\ 8.28]\  \text{(LSND)} \\ 
\lbrack -4.2,\ 6.6]\  \  \text{(TEXONO)}%
\end{array}%
$ & $%
\begin{array}{c}
\leq 1.5 \\ 
\text{(Reactor)}%
\end{array}%
\ $ & $-$ \\ \hline
$\nu _{\mu }($Others$)$ & \multicolumn{1}{|l|}{$%
\begin{array}{c}
\leq 5.8\  \text{(Borexino)\ } \\ 
\leq 68\  \text{(LSND)\ } \\ 
\leq 74\  \text{(LAMPF)}[7]%
\end{array}%
$} & \multicolumn{1}{|l|}{$%
\begin{array}{c}
\lbrack -9,\ 31]\  \text{(Solar)} \\ 
\leq 1.2\  \text{(CHARM-II)} \\ 
\lbrack -4.2,\ 0.48]\  \text{(TEXONO)}%
\end{array}%
$} & $-$ & $-$ \\ \hline
$\nu _{\tau }($Others$)$ & \multicolumn{1}{|l|}{$%
\begin{array}{l}
\leq 5.8\  \text{(Borexino)} \\ 
\multicolumn{1}{c}{\leq 3.9\times 10^{4}\  \text{(DONUT)}}%
\end{array}%
$} & $%
\begin{array}{l}
\lbrack -9,\ 31]\  \text{(Solar)}%
\end{array}%
$ & $%
\begin{array}{c}
\leq 3\times 10^{8}\  \\ 
\text{(Beam dump)}%
\end{array}%
$ & $-$ \\ \hline \hline
\end{tabular}%
\\[0pt]
\end{center}
\caption{90\% C.L. bounds on neutrino magnetic moment, charge radius,
 millicharge and anapole moment from XENON1T and other
 laboratory experiments. For comparison with astrophysical
 constraints see ref. \cite{ Giunti:2015gga} and for COHERENT see refs. \cite{Cadeddu:2020lky,  Khan:2019cvi}. Apart from Borexino \cite{Borexino:2017fbd} and
 solar \cite{Khan:2017djo}, bounds from all other experiments
 were taken from tables of ref. \cite{ Giunti:2015gga}.}
\label{tabel1}
\end{table*}

\begin{table*}[t]
\begin{center}
\begin{tabular}{c|c|c|c|c}
\hline \hline
Coupling & XENON1T & GEMMA & Borexino & TEXONO \\ \hline
$g_{V/A}(\times 10^{-7})$ & $\lesssim 3.5$ & $\lesssim 4.0$ & $%
\lesssim 10.8$ & $\lesssim 15$ \\ 
$g_{S}\ (\times 10^{-6})$ & $\lesssim 0.9$ & $-$ & $-$ & $-$ \\ 
$g_{P}\ (\times 10^{-6})$ &  $\lesssim 2.0$ & $-$ & $-$ & $-$ \\ \hline \hline
\end{tabular}%
\end{center}
\caption{90\% C.L. bounds at 10 keV mass of new light mediators of the vector, axial-vector, scalar and pseudoscalar couplings from XENON1T data (this work) and from other laboratory experiments (taken from ref. \cite{Harnik:2012ni}). For the full parameter space of all interaction for the mass ranges (0-100)keV from XENON1T data, see fig. 4}
\end{table*}

\section{Conclusions}
We did a constrained spectral analysis weighted by the $1\sigma$ experimental uncertainties to explain the excess observed in the electron recoil energy spectrum in XENON1T detector with neutrino nonstandard interactions. We have considered the popular new physics scenarios arising from the neutrino standard neutrino interaction at the low energy recoils. To this aim we have considered the neutrino
magnetic moment,  charge radius, millicharge, anapole moment and light vector/axial-vector scalar and pseudo-scalar mediators. The constraints on the fits were chosen in the ranges allowed by the other laboratory experiments as shown Table I. All the spectral fit results are shown in fig. 1 and 2.\par
With the help of this we have attempted to explain the observed excess using 1 parameter at-a-time fits. Realizing that a multi-parameter will explain excess more effectively, we have indicated here that even single parameter fits also show outstanding sensitivity in  some cases in low energy excess region. Our analysis shows that millicharge neutrino and vector mediators outperform to explain the observed excess. We leave multi-parameter fitting analysis for the follow up.\par
The key message of this analysis is the confirmation that any new physics sensitivity strongly depends on the order of the inverse momentum transfer as can be seen in the cases of millicharge and vector mediators. For the light mediators, due to a large allowed free parameter space of masses and couplings constants the excess be explained with arbitrary but reasonable choices already determined by other experiments. In this case we find that the whole spectrum, but the excess region in particular, is sensitive to variation in a very narrower window of the parameter boundaries considered here.\par
We have also derived bounds on all new physics parameters considered in this work and the results are shown in fig. 3 and 4 and in Table I. Some of them are in \textit{tension} with the constraints from COHERENT (see Table II and IV of ref. \cite{Cadeddu:2020lky}) and other experiments (see Table I here).
The bounds on neutrino magnetic moment obtained in ref. \cite{Cadeddu:2020lky} are too large to explain the excess here, while those on charge radius are too tight to explain this excess. This implies that more stringent constraints were expected on neutrino magnetic moment but relatively weaker on charge radius are expected from the XENON1T data. This is what we have obtained in section $\text{III-B}$ and Table I. 
\par We present constraints on the vector and scalar couplings obtained from this study at 10 keV mediator masses in Table 1. At the same mass value we also show bounds from the other laboratory experiments, GEMMA, Borexino and Texono \cite{Harnik:2012ni} for the vector couplings. The full $\Delta\chi^{2}$ distributions for the other benchmark values of masses are shown in fig. 4. Comparison with other bounds show that there is a factor of 3-4 improvement from Borexino and TEXONO while the GEMMA bound is slightly improved. It is important to note that in spite of the improvement by XENON1T of the vector couplings, they still cannot  compete with the direct bound from astrophysical bounds. However, this is a starting from the direct detection experiments and upgrade of XENON1T or the other twins experiments XENONnT, LZ, Darkside-20k and DARWIN \cite{Aalbers:2020gsn, Aalseth:2017fik, Aalbers:2016jon} are likely to significantly improve these constraint or our understanding of the neutrino interactions more directly. Notice that the constraints on the general type axial-vector, scalar and pseudoscalar are first time reported in this study up-to the best of our knowledge.    
\par
We find that within $1\sigma $ experimental
uncertainty the excess region of $(1-7)\ $keV is better explained in the
range$\ (2-4)\times 10^{-11}\mu _{B}$ for magnetic moment,
$(1.7-2.3)\times 10^{-12}e$ for the millicharge, and $(10-100)\ $keV masses of light mediators with couplings of $ 5\times 10^{-7}$ for vector/axial-vector, and $1.2\times 10^{-6}$ and $4\times 10^{-6}$ for scalar and pseudo-scalar mediators, respectively. However, as shown directly from the spectral fit in fig. 1 for NCR and from the two parameter fitting in fig. 5, the neutrino millicharge is the leading candidate to outstandingly explain the observed excess. This avenue requires further a careful investigation. The constraints on neutrino millicharge neutrinos derived in this work, however, are around 2 order of magnitude weaker than those from astrophysical observations \cite{Raffelt:1999gv,Davidson:2000hf}.
\par We conclude that neutrino nonstandard interactions could be the top candidate to explain the observed excess by XENON1T given the fact that neutrinos are massive and could have new interactions beyond the standard model unless this possibility is excluded by other experiments \cite{Aalbers:2020gsn, Aalseth:2017fik, Aalbers:2016jon, Coloma:2020voz}.



\begin{acknowledgments}
The author thanks Werner Rodejohann for useful discussions. This work is financially supported by Alexander von Humboldt Foundation under the postdoctoral fellowship program.
\end{acknowledgments}

\bibliography{biblio}

\end{document}